\documentclass[12pt]{article}

\usepackage{setspace,amsmath,graphicx,epsfig, bm, color,calc,graphics, times, diagbox}
\usepackage{natbib, amsthm}

\newcommand{\etal}{et al.}

\def\bZ{\mathbf Z}
\def\Var{{\rm Var}\,}

\def\E{{\rm E}\,}
\def\boxit#1{\vbox{\hrule\hbox{\vrule\kern6pt\vbox{\kern6pt#1\kern6pt}\kern6pt\vrule}\hrule}}

\newtheorem{theorem}{Theorem}
\newtheorem{lemma}{Lemma}
\newtheorem{corollary}{Corollary}
\newtheorem{remark}{Remark}

\title{Functional and Parametric Estimation in a Semi- and Nonparametric Model with Application to Mass-Spectrometry Data}

\author{Weiping Ma$^{1}$, 
Yang Feng$^{2}$,  
Kani Chen$^{3}$, and
Zhiliang Ying$^{2}$\\
\small{$^{1}$School of Public Health,
Fudan University, Shanghai 200433, China, }\\
\small{$^{2}$Department of Statistics, Columbia University, New York, NY 10027, USA. }\\
\small{$^{3}$Department of Mathematics,
Hong Kong University of Science and Technology, Kowloon,
Hong Kong.}}
\date{}
\begin{document}

\maketitle
\begin{abstract}

Motivated by modeling and analysis of mass-spectrometry data, a semi- and nonparametric model is proposed that consists of a linear parametric component for individual location and scale and a nonparametric regression function for the common shape. A multi-step approach is developed that simultaneously estimates the parametric components and the nonparametric function. Under certain regularity conditions, it is shown that the resulting estimators is consistent and asymptotic normal
for the parametric part and achieve the optimal rate of convergence for the nonparametric part when the bandwidth is suitably chosen. Simulation results are presented to demonstrate the effectiveness and finite-sample performance of the method.  The method is also applied to a SELDI-TOF mass spectrometry data set from a study of liver cancer patients.


\end{abstract}
%
%

\section{Introduction}
We are concerned with the following semi- and nonparametric regression model

\begin{equation}\label{model}
y_{it}=\alpha_i+\beta_i m(x_{it})+\sigma_i(x_{it})\epsilon_{it},
\end{equation}
where $y_{it}$ is the observed response from $i$-th individual $(i = 1,\ldots,n)$ at time $t$ for $(t = 1,\ldots,T)$, $x_{it}$ is the corresponding explanatory variable,   $\alpha_i$ and $\beta_i$ are individual-specific location and scale parameters and $m(\cdot)$ is a baseline intensity function. Here, $E \epsilon_{it} =0 $, $\Var(\epsilon_{it}) =1$, and $\epsilon_{it}$ and $x_{it}$ are independent. Of interest here is the simultaneous estimation of  $\alpha_i$, $\beta_i$ and $m(\cdot)$. We shall assume throughout the paper that $\epsilon_{it}$  $(i = 1,\ldots, n ; t = 1,\ldots,T)$ are independent and identically distributed (i.i.d.) with an unknown distribution function, though most results only require that the errors be independent with zero mean.


Model (\ref{model}) is motivated by analyzing the data generated from mass spectrometer (MS), which is a powerful tool for the separation and large-scale detection of proteins present in a complex biological mixture. Figure 1 is an illustration of MS spectra, which can reveal proteomic patterns or features that might be related to specific characteristic of biological samples. They can also be used for prognosis and for monitoring disease progression,  evaluating treatment or suggesting intervention. Two popular mass spectrometers are SELDI-TOF (surface enhanced laser desorption/ionization time-of-fight) and MALDI-TOF (matrix assisted laser desorption and ionization time-of-flight).  The abundance of the protein fragments from a biological sample (such as serum) and their time of flight through a tunnel under certain electrical pressure can be measured by this procedure. The
$y$-axis of a spectrum is the intensity (relative abundance) of
protein/peptide, and the $x$-axis is the mass-to-charge ratio (m/z value) which can be calculated using time,  length of flight, and the voltage applied. It is known that the SELDI intensity
measures have errors up to 50\% and that the $m/z$ may shift
its value by up to 0.1\%--0.2\% \citep{Yasui}.  Generally speaking, many pre-processing steps need to be done before the MS data can be analyzed. Some of the most important steps are noise filtering, baseline correction, alignment, normalization, etc. See, e.g., \cite{Guilhaus, Banks03, Baggerly03,Baggerly04, Diamandis, Feng-2009}.  We refer readers to \cite{Roy-2011} for an extensive review about the recent advances in mass-spectrometry data analysis. Here, we assume all the pre-processing steps have already been taken.

\begin{figure}[t]
\caption{Illustration of MS spectra.}
    \begin{center}
      \includegraphics[scale=0.5]{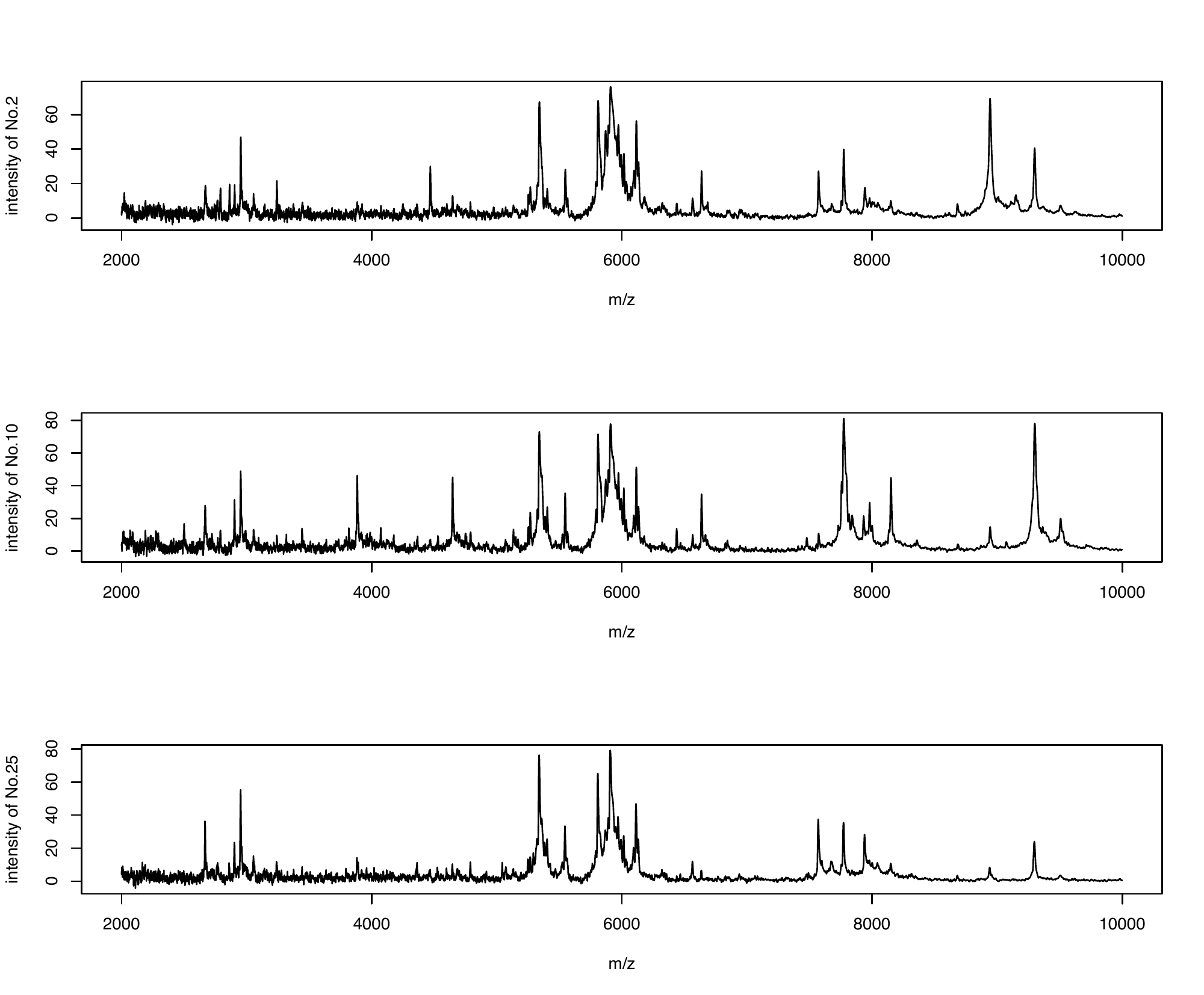}
    \end{center}
\end{figure}

In model (\ref{model}), $m(\cdot)$ represents the common shape for all individuals while $\alpha_i$ and $\beta_i$ represents the location and scale parameters for the $i$-th individual, respectively. Because $m(\cdot)$ is unspecified, model \eqref{model} may be viewed as a semiparameteric model. However, it differs from the usual semi-parametric models in that for model (\ref{model}), both the parametric and nonparametric components are of primary interest, while in a typical semiparametric setting, the nonparametric component is often viewed as a nuisance parameter. Model \eqref{model} contains many commonly encountered regression models as special cases. If all the parametric coefficients $\alpha_i$ and $\beta_i$ are known, model (\ref{model})  reduces to the classical nonparametric regression. On the other hand, if the function $m(\cdot)$ is known, then it reduces to the classical linear regression model with each subject having its own regression line. For the present case of $\alpha_i$, $\beta_i$ and function $m(\cdot)$ being unknown, the parameters are identifiable only up to a common location-scale change. Thus we assume, without loss of generality, that $\alpha_1=0$ and $\beta_1=1$. It is also clear that for $\alpha_i$, $\beta_i$ and $m(\cdot)$ to be consistently estimable, we need to require that both $n$ and $T$ go to $\infty$.

There is an extensive literature on semiparametric and nonparametric regression. For semiparametric regression,  \cite{Begun-1983} derived semiparametric information bound while \cite{Robinson-1988} developed a general approach to constructing $\sqrt{n}$-consistent estimation for the parametric component. We refer to \cite{Bickel-1998} and \cite{Ruppert-2003} for detailed discussions on the subject. For nonparametric regression, kernel and local polynomial smoothing methods are commonly used \citep{Rosenblatt-1956,Stone-1977,Stone-1982,Fan-1993}. In particular, local polynomial smoothing has many attractive properties including the automatic boundary correction. We refer to \citet{Fan-Gijbels-1996} and \cite{Hardle-2004} for comprehensive treatment of the subject.

The existing methods for dealing with nonparametric and semiparametric problems are not directly applicable to model \eqref{model}. This is due to the mixing of the finite dimensional parameters and the nonparametric component. A natural way to handle such a situation is to de-link the two aspects of the estimation through a two-step approach. In this paper, we propose an efficient iterative procedure, alternating between estimation of  the parametric component and the nonparametric component. We show that the proposed approach leads to consistent estimators for both the finite-dimensional parameter and the nonparametric function. We also establish asymptotic normality for parametric estimator and convergence rate for the nonparametric estimation that is then used for optimal bandwidth selection.


\section{Main Results}\label{sec::main.results}
In this section, we develop a multi-step approach to estimating both the finite-dimensional parameters $\alpha_i$ and $\beta_i$ and the nonparametric baseline intensity $m(\cdot)$. Our approach is an iterative procedure which alternates between estimation of $\alpha_i$ and $\beta_i$ and that of $m(\cdot)$. We show that under reasonable conditions, the estimation for the parametric component is consistent and asymptotically normal when the bandwidth selection are done appropriately. The estimation of the nonparametric component can also attain the optimal rate of convergence.

\subsection{A multi-step estimation method}\label{sec::estimation-method}
Recall that if $\alpha_i$ and $\beta_i$ were known, the problem would reduce to the standard nonparametric regression setting; on the other hand, if $m(\cdot)$ were known, it would reduce to the simple linear regression for each $i$.
For the nonparametric regression, we can apply the local linear regression with the weights $K_h(\cdot)=K(\cdot/h)/h$ for suitably chosen kernel function $K$ and bandwidth $h$.  For the simple linear regression, the least squares estimation may be applied.

Not all parameters in model \eqref{model} are identifiable as $\alpha_i$, $\beta_i$ and $m(\cdot)$ are confounded. To ensure identifiability, we shall set $\alpha_1=0$ and $\beta_1=1$. Thus, for $i=1$, \eqref{model} becomes a standard nonparametric regression problem, from which an initial estimator of $m(\cdot)$ can be derived. Replacing $m(\cdot)$ in \eqref{model} by the initial estimator, we can apply the least squares method to get estimators of $\alpha_i$, $\beta_i$ for $i\ge 2$, which, together with $\alpha_1=0$ and $\beta_1=1$ and local polynomial smoothing, can then be used to get an updated estimator of $m(\cdot)$. This iterative estimation procedure is described as  follows.

\begin{itemize}
  \item[(a)]  Set $\alpha_1=0$ and $\beta_1=1$, so that $y_{1t}=m(x_{1t})+\sigma_1(x_{1t})\epsilon_{1t}$, $t=1,\ldots,T$.
Apply local linear regression to $(x_{1t},y_{1t}), t=1,\ldots, T$, to get initial estimator of $m(\cdot)$
\begin{align}\label{eq::initial_estimate_tilde_m}
\tilde m(x)=\frac{\sum_{t=1}^T\omega_{1t}(x)y_{1t}}{\sum_{t=1}^T\omega_{1t}(x)},
\end{align}
where $\omega_{1t}(x)=K_h(x_{1t}-x)(S_{T,2}-(x_{1t}-x)S_{T,1})$ and
   \begin{align}\label{eq::S_Tk}
     S_{T,k}=\sum_{t=1}^TK_h(x_{1t}-x)(x_{1t}-x)^k\mbox{, for }k=1, 2.
   \end{align}

\item [(b)] With $m(\cdot)$ being replaced by $\tilde m(\cdot)$ as the true function, $\alpha_i$, $\beta_i$, $i=2, \dots, n$ can be estimated by the least squares method, i.e.   \begin{align}
 \hat\beta_i=\frac{\sum_{t=1}^T[\tilde m(x_{it})-\bar{\tilde m}(x_{i\cdot})]y_{it}}{\sum_{t=1}^T[\tilde m(x_{it})-\bar{\tilde m}(x_{i\cdot})]^2},
  \end{align}
    \begin{align}
      \hat\alpha_i=\bar y_{i\cdot}-\hat\beta_i\bar{\tilde m}(x_{i\cdot})=\bar y_{i\cdot}-\frac{\sum_{t=1}^T(\tilde m(x_{it})-\bar{\tilde m}(x_{i\cdot}))y_{it}}{\sum_{t=1}^T(\tilde m(x_{it})-\bar{\tilde m}(x_{i\cdot}))^2}\bar{\tilde m}(x_{i\cdot}),
    \end{align}
    where $$\bar y_{i\cdot}=\frac{1}{T}\sum_{t=1}^Ty_{it}, \mbox{ and } \bar{\tilde m}(x_{i\cdot})=\frac{1}{T}\sum_{t=1}^T\tilde m(x_{it}).$$
\item [(c)]With the estimates $\hat\alpha_i$ and $\hat\beta_i$, we can update the estimation of $m(\cdot)$ viewing $\hat\alpha_i$ and $\hat\beta_i$ as true values. Specifically, we apply the local linear regression with the same kernel function $K(\cdot)$  to get an updated estimator of $m(\cdot)$,
    \begin{align}\label{eq::hat_m}
   \hat m(x)=\frac{\sum_{i=1}^n\sum_{t=1}^T\omega_{it}^*(x)y_{it}^*}{\sum_{i=1}^n\sum_{t=1}^T\omega_{it}^*(x)},
    \end{align}
where $y_{it}^*=(y_{it}-\hat\alpha_i)/\hat\beta_i$,
    \begin{align}\label{eq::omegastar}
       \omega_{it}^*(x)
    =&\hat\beta_i^2K_{h^*}(x_{it}-x)\left[\sum_{i=1}^n\hat\beta_i^2S_{T,2}^{*(i)}-(x_{it}-x)\sum_{i=1}^n\hat\beta_i^2S_{T,1}^{*(i)}\right]
    \end{align}
    and
    \begin{align} \label{eq::S_Tkstar}
S_{T,k}^{*(i)}=\sum_{t=1}^TK_{h^*}(x_{it}-x)(x_{it}-x)^k\mbox{, for }k=1, 2.
    \end{align}
Note that the bandwidth for this step, $h^*$, may be chosen differently from $h$ in order to achieve better convergence rate. The optimal choices for $h$ and $h^*$ will become clear in the next subsection where large sample properties are studied.
\item [(d)] Repeat steps (b) and (c) until both the parametric and the nonparametric estimators converge.

\end{itemize}

Our limited numerical experiences indicate that the final estimator is not sensitive to the initial estimate. However, as a safe guard, we may start the algorithm with different initial estimates by choosing different individuals as the baseline intensity. In step (c), the $\hat\beta_i$ is in the denominator, which, when close to $0$, may cause instability. Thus, in practice, we can add a small constant to the denominator to make it stable, though we have not encountered this problem.

The iterative process often converges very quickly. In addition, our asymptotic analysis in the next subsection shows that no iteration is needed to reach the optimal convergence rate for the estimate of both parametric and nonparametric components when the bandwidths of each step are properly chosen. Therefore, we may stop after step (c) to save computation time for large problems.

\subsection{Large Sample Properties}\label{sec::large:sample:properties}
In this section, we study the large sample properties of the estimates for $m(\cdot)$, $\alpha_i$ and $\beta_i$. By large sample, we mean that both $n$ and $T$ are large. However, the size of $n$ and that of $T$ can be different. Indeed, for MS data, $T$ is typically much larger than $n$. The optimal bandwidth selection in the nonparametric estimation will be determined by asymptotic expansions to achieve optimal rate of convergence. We will also investigate whether or not the accuracy of $\hat\alpha_i$ and $\hat\beta_i$ may  affect the rate of convergence for the estimation of $m(\cdot)$.

The following conditions will be needed to establish the asymptotic theory.
\begin{itemize}

\item[C1.] The baseline intensity $m(\cdot)$ is continuous and has a bounded second order derivative.
\item[C2.] There exist constants $\alpha>0$ and $\delta>0$, such that the marginal density $f(\cdot)$ of $x_{it}$ satisfies $f(x)>\delta$, and $|f(x)-f(y)|\leq c|x-y|^\alpha$ for any $x$ and $y$ in the support of $f(\cdot)$.
\item[C3.] The conditional variance $\sigma_i^2(x)=\Var(y_{it}|x_{it}=x)$ is bounded and continuous in $x$, where $i=1,\ldots,n$ and $t=1,\ldots, T$.
\item[C4.] The kernel $K(\cdot)$ is a symmetric probability density function with bounded support. Hence $K(\cdot)$ has the properties:
  $\int^{\infty}_{-\infty}K(u)du=1,\int^{\infty}_{-\infty}uK(u)du=0,\int^{\infty}_{-\infty}u^2K(u)du\neq0$ and bounded. Without loss of generality, we could further assume the support of $K(\cdot)$ lies in the interval $[-1,+1]$.

\end{itemize}
Condition C1 is a standard condition for nonparametric estimation. Condition C2 requires that the density of $x_{it}$ is bounded away from 0, which may be a strong assumption in general but reasonable for mass spectrometry data as $x_{it}$ are approximately uniformly distributed on the support. In addition, the density is assumed to satisfy a Lipschitz condition.   Condition C3 allows for heteroscedasticity while restricting the variances to be bounded. Condition C4 is a standard  condition for kernel function used in the local linear regression.

The moments of $K$ and $K^2$ are denoted respectively by  $\mu_l=\int^{\infty}_{-\infty}u^lK(u)du$ and $\nu_l=\int^{\infty}_{-\infty}u^lK^2(u)du$ for $l\geq 0$.
\begin{lemma}\label{lm1}
Suppose that Conditions C1-C4 are satisfied. Then for $\tilde m(\cdot)$ defined by \eqref{eq::initial_estimate_tilde_m}, we have, as $h\rightarrow 0$ and $Th\rightarrow\infty$,
\begin{eqnarray}\label{eq-lm1}
\tilde m(x)=m(x)+\frac{1}{2}m''(x)\mu_2h^2+o(h^2)+U_1(x),
\end{eqnarray}
where $U_1(x)=(\sum_{t=1}^T\omega_{1s}(x)\sigma_1(x_{1s})\epsilon_{1s})/(\sum_{t=1}^T\omega_{1s}(x))$.

\end{lemma}

Lemma \ref{lm1} allows us to derive the asymptotic bias, variance and mean squared error for the estimator $\tilde m(\cdot)$. This is summarized in the following corollary.

\begin{corollary}\label{co1}
Let $\mathbf X$ denote all the observed covariates $\{x_{it},i=1,\ldots,n,t=1,\ldots,T\}$. Under Conditions C1-C4, the bias, variance and mean squared error of $\tilde m(x)$ conditional on $\mathbf X$ have the following expressions.
\begin{eqnarray*}
&&\E(\tilde m(x)-m(x)\big|\mathbf X)=\frac{1}{2}m''(x)\mu_2h^2+o(h^2),\\
&&\Var(\tilde m(x)\big|\mathbf X)=\frac{1}{Th}[f(x)]^{-1}\sigma_1^2(x)\nu_0+o\Big(\frac{1}{Th}\Big),\\
&&\E[\{\tilde m(x)-m(x)\}^2\big|\mathbf X]=\frac{1}{4}(m''(x)\mu_2)^2h^4+\frac{1}{Th}[f(x)]^{-1}\sigma_1^2(x)\nu_0+o\Big(h^4+\frac{1}{Th}\Big).
\end{eqnarray*}

\end{corollary}

It is clear from the above expansions that in order to minimize the mean squared error of $\tilde m(x)$,  the bandwidth $h$ should be chosen to be of order $T^{-1/5}$. However, we will show later that this is not necessarily optimal for our final estimator $\hat m(\cdot)$.

For estimation of scale parameters $\beta_i$, we can apply Lemma \ref{lm1} together with the Taylor expansion to derive asymptotic bias and variance. In particular, we have the following theorem.

\begin{theorem}\label{th1}
Suppose that Conditions C1-C4 are satisfied and that $h\rightarrow 0$ is chosen so that $Th\rightarrow\infty$. Then the following expansions hold for $i\geq 2$.

\begin{eqnarray}
&&\E(\hat\beta_i-\beta_i\big|\mathbf X)=-\beta_i\Big(h^2P_i+\frac{1}{Th}Q_i\Big)+o\Big(h^2+\frac{1}{Th}\Big),\label{eq::beta-bias}\\
&&\Var(\hat\beta_i\big|\mathbf X)=\frac{\sum_{t=1}^TW_{it}^2\sigma_i^2(x_{it})}{(\sum_{t=1}^TW_{it}^2)^2}+\beta_i^2 \frac{\sum_{t=1}^TW_{1t}^2\sigma_i^2(x_{1t})}{(\sum_{t=1}^TW_{it}^2)^2}+o\Big(\frac{1}{T}\Big),\label{eq::beta-var}
\end{eqnarray}
where $W_{it}=m(x_{it})-\bar m(x_{i\cdot}), \bar m(x_{i\cdot})=T^{-1}\sum_{t=1}^Tm(x_{it}),$
$$P_i=\frac{\mu_2}{2}\frac{\sum_{t=1}^TW_{it}m''(x_{it})}{\sum_{t=1}^TW_{it}^2},Q_i=\frac{\nu_0\sum_{t=1}^Tf^{-1}(x_{1t})\sigma_i^2(x_{1t})}{\sum_{t=1}^TW_{it}^2}.
$$
\end{theorem}

\begin{remark}\label{rm2}
The asymptotic bias and variance of parameter estimator $\hat\alpha_i$ can be similarly derived. In fact, they can be inferred from the bias and variance of $\hat\beta_i$ through its linear relationship with $\hat\beta_i$, thus having the same order as those of $\hat\beta_i$ in \eqref{eq::beta-bias} and \eqref{eq::beta-var}.
\end{remark}
\begin{remark}\label{rm3}
 The bias of $\hat\beta_i$ is of the order $h^2+(Th)^{-1}$ and the variance is of the order $T^{-1}$. To obtain the $\sqrt{T}$-consistency for $\hat\beta_i$, i.e. $\sqrt T(\hat\beta_i-\beta_i)=O_p(1)$, the order of bias should be $O(T^{-1/2})$. This is achieved by choosing $h$ to be between $T^{-1/2}$ and $T^{-1/4}$.
\end{remark}

From the asymptotic expansion for the mean and variance of the initial functional estimator $\tilde m(\cdot)$ and parameter estimator $\hat\beta_i$, we can obtain the asymptotic expansions for the bias and variance of the subsequent estimator of the baseline intensity, $\hat m(\cdot)$.

\begin{theorem}\label{th2}
Suppose that Conditions C1-C4 are satisfied. Suppose also that $h$ for $\tilde m(\cdot)$ and $h^*$ for $\hat m(\cdot)$ are chosen so that  $h\rightarrow 0$, $h^*\rightarrow 0$, $Th\rightarrow\infty$, and $nTh^*\rightarrow\infty$. Then the following expansions hold:
\begin{align*}
&\E(\hat m(x)-m(x)\big|\mathbf X)\\
&\quad=  \left(\sum_{i=2}^n\frac{\beta_i^2( h^2P_i+(Th)^{-1}Q_i) }{\sum_{i=1}^n\beta_i^2}\right)  m(x)-\sum_{i=2}^n\frac{\beta_i^2(  h^2R_i+(Th)^{-1}m(x_{i\cdot})Q_i )}{\sum_{i=1}^n\beta_i^2}\\
&\quad\quad+\frac{m''(x)\mu_2}{2}h^{*2}+o\Big(h^2+\frac{1}{Th}+h^{*2}\Big),
\end{align*}
\begin{align*}
\Var(\hat m(x)\big|\mathbf X)&=\frac{1}{(\sum_{i=1}^n\beta_i^2)^2}\sum_{t=1}^T\left(\sum_{i=2}^n\beta_i^2\Big[\frac{1}{T}+Z_{it}
\Big]\right)^2\sigma_1^2(x_{1t})\\&+\frac{\nu_0\sum_{i=2}^n\beta_i^2f^{-1}(x)\sigma_i^2(x_{it})}{Th^*(\sum_{i=1}^n\beta_i^2)^2}+o\Big(\frac{1}{T}+\frac{1}{nTh^*}\Big).\\
\end{align*}
where $P_i, Q_i, W_{it}$ are the same as those in Theorem \ref{th1}, and $R_i=\bar m(x_{i\cdot})P_i-2^{-1}\mu_2\bar{ m''}(x_{i\cdot})$, $Z_{it}=(\sum_{s=1}^TW_{is}^2)^{-1}(m(x)-\bar m(x_{i\cdot}))W_{1t}$.

\end{theorem}

%
%

In the ideal case when the location-scale parameters are known, the bias and variance of the local linear estimator of baseline intensity $m(\cdot)$ should be of the order $O(h^{*2})$  and $O(\frac{1}{nTh^*})$. And the optimal bandwidth in this ideal case should be of order $(nT)^{-\frac{1}{5}}$. Therefore the bias and variance of the nonparametric estimator are $O((nT)^{-\frac{2}{5}})$ and $O((nT)^{-\frac{4}{5}})$, respectively. In addition, the mean squared error is of order $O((nT)^{-\frac{4}{5}})$. Interestingly, by choosing the bandwidths $h$ and $h^*$ separately, we can achieve this optimal rate of convergence for the baseline intensity estimator $\hat m(\cdot)$ through the proposed multi-step estimation procedure when the orders of $n$ and $T$ satisfy certain requirement. Notice that the parametric components will have the optimal $\sqrt{T}$ convergence rate simultaneously.  The conclusions are summarized in the following theorem.


\begin{theorem}\label{th3}
Suppose that Conditions  C1-C4 are satisfied. The optimal parametric convergence rate of location-scale estimators can be attained by setting $h$ to be of order $T^{-\frac{1}{3}}$; the optimal nonparametric convergence rate of the baseline intensity estimator $\hat m(\cdot)$ can be attained by setting $h^*$ to be of order $(nT)^{-\frac{1}{5}}$ and $h$ of order $T^{-\frac{1}{3}}$, when $T\rightarrow\infty$, $n\rightarrow\infty$, and $n = O(T^{1/4})$.
\end{theorem}

\begin{remark}\label{rm6}
It is clear from Theorem \ref{th3} that if the requirement $n = O(T^{1/4})$ is not satisfied, then the nonparametric estimator $\hat m(\cdot)$ will not achieve the optimal rate of convergence at any choice of the bandwidths. However, the choice of $h$ and $h^*$ is optimal even if $n=O(T^{1/4})$ does not hold.
\end{remark}

\begin{theorem}\label{th4}
Suppose that Conditions C1-C4 are satisfied. In addition, assume $E[m^2(x_{it})(\sigma_i^2(x_{it})+1)]<\infty$ and $E[m^2(x_{it})]>0$ for all $i = 1,...,n$ and $t=1,\ldots,T$.
If we restrict the order of $h$ to lie between $T^{-\frac{1}{2}}$ and $T^{-\frac{1}{4}}$, $\hat\beta_i$ is asymptotic normal:
\begin{align}\label{eq::betainormal}
\sqrt T(\hat\beta_i-\beta)\rightarrow N(0, \sigma_i^{*2}),
\end{align}
where
\begin{align*}
\sigma_i^{*2}&=\lim_{T\rightarrow\infty}\left(\frac{T^{-1}\sum_{t=1}^TW_{it}^2\sigma_i^2(x_{it})}{(T^{-1}\sum_{t=1}^TW_{it}^2)^2}+\beta_i^2 \frac{T^{-1}\sum_{t=1}^TW_{1t}^2\sigma_1^2(x_{1t})}{(T^{-1}\sum_{t=1}^TW_{it}^2)^2}\right).
\end{align*}

\end{theorem}
Here, if we assume $\sigma_i^2(\cdot)$ to be a constant for each subject $i$, then its value can be consistently estimated by the plug-in estimator \begin{align}\hat\sigma_i^2=T^{-1}\sum_{t=1}^T(y_{it}-\hat\alpha_i-\hat\beta_i\hat m(x_{it}))^2,
 \end{align}
where $\hat\alpha_1=0$ and $\hat\beta_1=1.$

From \eqref{eq::betainormal}, the asymptotic variance of  $\hat\beta_i$ is of order $O(T^{-1})$, provided that the order of the bandwidth $h$ is properly chosen. Since the asymptotic expansion for $\hat\beta_i$ does not involves the choice of $h^*$, the specific choice of different $h^*$ will not affect the order of the asymptotic variance of $\hat\beta_i$.


\subsection{Bandwidth Selection}\label{sec::bandwidth.selection}
In Section \ref{sec::large:sample:properties}, we studied how the choice of bandwidths $h$ and $h^*$ may affect the asymptotic properties of the estimators. However, in practice, we need a data-driven approach to choosing the bandwidths. Our suggestion on this is to use a $K$-fold cross-validation bandwidth selection rule.

First, we divide the $n$ individuals into $K$ groups $\bZ_1,\bZ_2,\ldots,\bZ_K$ randomly. Here, $\bZ_k$ is the $k$-th test set, and the $k$-th training set is $\mathbf{Z}_{-k}=\{\{1,\ldots,n\}\backslash \bZ_k\}$. We estimate the baseline curve $m(\cdot)$ using the observations in the training set $\mathbf{Z}_{-k}$ and denote the estimator as $\hat m(\mathbf{Z}_{-k},h,h^*)$, where $h$ and $h^*$ are the  bandwidths of the two nonparametric regression steps for $\tilde m(\cdot)$ and $\hat m(\cdot)$, respectively.   Recall that at the beginning of the multi-step estimation procedure, we fix the first observation as the baseline to solve the identifiability issue. In the case of cross-validation, for each split, the baseline will corresponds to the first observation inside $\mathbf{Z}_{-k}$, which is different for different $k$. We circumvent the problem of comparing different baseline estimates by using them to predict the test data  in $\mathbf{Z}_k$, i.e., after obtaining the estimator of baseline curve from $\mathbf{Z}_{-k}$. We then regress it on the data in $Z_k$, and compute the mean squared prediction  error (MSPE).
\begin{equation}
{MSPE}(\mathbf{Z}_{-k},h,h^*)=\frac{1}{T}\sum_{i\in \mathbf{Z}_k}\sum_{t=1}^T[y_{it}-(\hat\alpha_{ki}+\hat\beta_{ki}\hat m_t(\mathbf{Z}_{-k},h,h^*))]^2,
\end{equation}
where $\hat\alpha_{ki}$ and $\hat\beta_{ki}$ are the estimated regression coefficients.
We repeat the calculation for $k= 1,\ldots,K$, and the optimal pair  $(\hat h,\hat h^*)$ is the one which minimizes the average MSPE, i.e.
\begin{equation}
(\hat h,\hat h^*)=\arg\min_{(h,h^*)}\frac{1}{K}\sum_{k=1}^K{MSPE}(\mathbf{Z}_{-k},h,h^*).
\end{equation}
The effectiveness of the cross-validation will be evaluated in Sections \ref{sec::realdata} and \ref{sec::simulation}.

\section{Application to Mass Spectrometry Data}\label{sec::realdata}
\begin{figure}[t]
\caption{Curves of 10 observations and the baseline estimate}\label{fig::real-estimates}
    \begin{center}
      \includegraphics[scale=0.5]{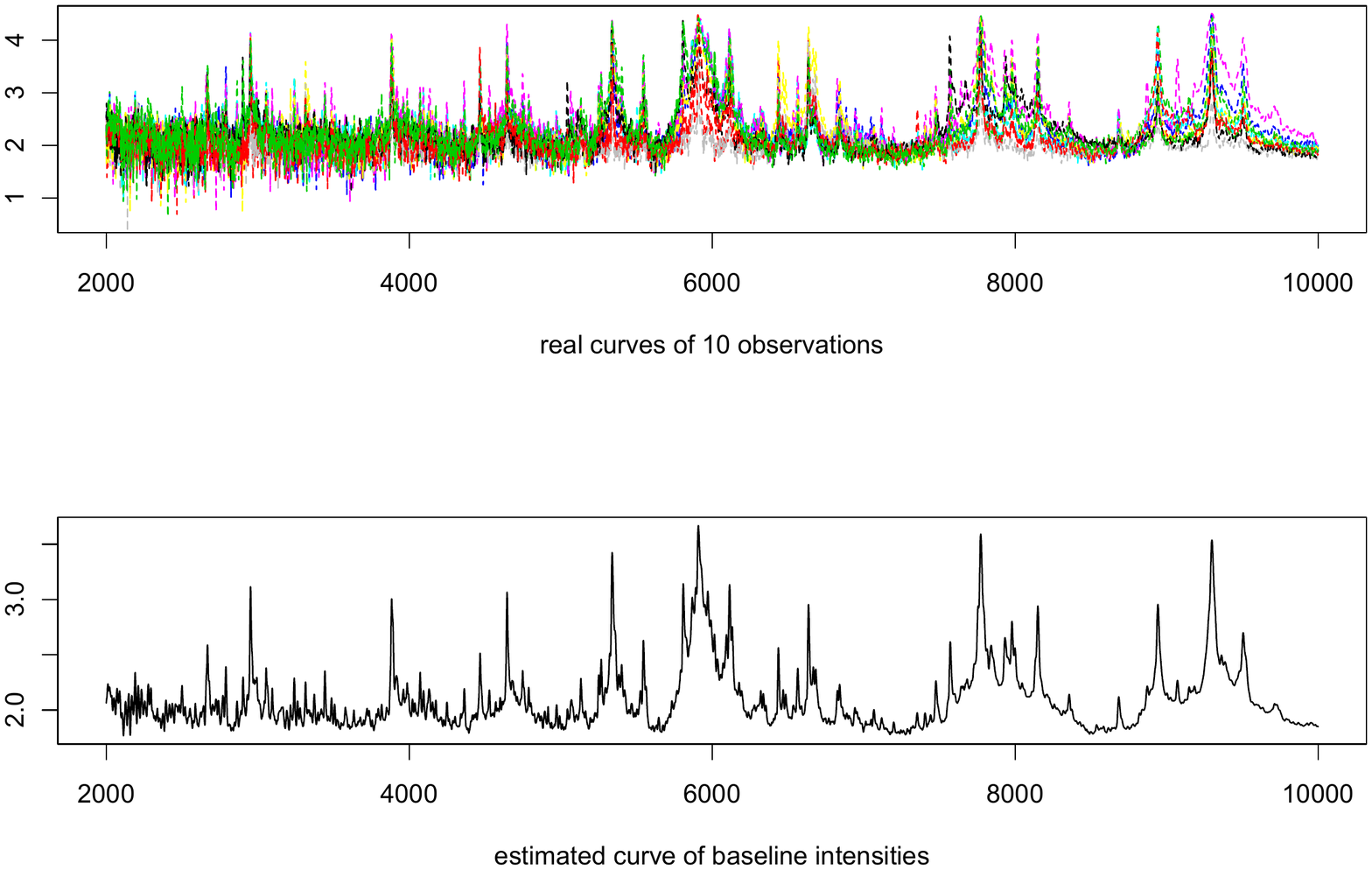}
    \end{center}

\end{figure}

We now apply the proposed multi-step method to a SELDI-TOF mass spectrometry data set from a study of 33 liver cancer patients conducted at Changzheng Hospital in Shanghai. For each patient, we extract the  $m/z$ values in the region 2000-10000 Da, which is believed to contain all the useful information. Figure \ref{fig::real-estimates} contains the curves of
10 randomly picked patients.

There are some noticeable features in the data. All curves appear to be continuous. They peak simultaneously around certain locations; at each location, curves have the same shape but with different heights. All those features are captured well by our model.

Since the observed values of $m/z$ for each person may fluctuate, we need to perform registration to make the analysis easier. Here, we use the observations from the first individual and set his/her $m/z$ values as the reference. Then we use the linear interpolation method to compute the intensities of all the other individuals at the reference $m/z$ locations. After that we get the preprocessed data which has the same $m/z$ values for each observation.

We use the cross-validation method described in Section \ref{sec::bandwidth.selection} to select the optimal bandwidths with $K=33$, i.e., leave-one-out cross validation.  We compute the MSPE at the grid of $h=2,4,6,\ldots,40$ and $h^*=2,4,6,\ldots,40$. Table \ref{tb::realdata:mspe} contains a portion of the result with $h=30,32,\ldots,40$ and $h^*=2,4,6,\ldots,20$.

\begin{table}[t] \centering 
\caption{MSE of the leave-one-out prediction of real data\label{tb::realdata:mspe}}
\begin{center}
\begin{tabular}{c|r|r|r|r|r|r}
\hline
\backslashbox{$h^*$}{$h$}&	30&	32&	34&	36&	38&	40\\
\hline
2&		1104.946&	1104.941&	1104.936&	1104.934&	 1104.931&	 1104.930\\
4&		1104.483&	1104.482&	\textbf{1104.481}&	1104.483&	 1104.484&	 1104.487\\
6&		1110.261&	1110.265&	1110.269&	1110.275&	 1110.281&	 1110.288\\
8&		1122.601&	1122.610&	1122.619&	1122.630&	 1122.640&	 1122.652\\
10&		1140.525&	1140.539&	1140.552&	1140.568&	 1140.582&	 1140.598\\
12&		1161.739&	1161.757&	1161.775&	1161.795&	 1161.813&	 1161.832\\
14&		1183.842&	1183.864&	1183.886&	1183.909&	 1183.931&	 1183.953\\
16&		1205.356&	1205.382&	1205.407&	1205.433&	 1205.458&	 1205.483\\
18&		1225.298&	1225.327&	1225.355&	1225.383&	 1225.411&	 1225.438\\
20&		1243.068&	1243.099&	1243.130&	1243.162&	 1243.191&	 1243.222\\
\hline

\end{tabular}
\end{center}
\end{table}

As we can see in Table \ref{tb::realdata:mspe}, the minimum MSE occurs at the location of $h=34$ and $h^*=4$, which agrees with our theory that $h$ and $h^*$ should not be chosen with the same rate for the purpose of estimating the nonparametric component.

Finally, we use the selected bandwidths to estimate the location and scale parameters as well as the nonparametric curve for the whole data set. The estimated parameters are reported in Table \ref{para2} and the baseline nonparametric curve estimation is shown in the lower part of Figure \ref{fig::real-estimates}. From Table \ref{para2}, we can see that each individual has very different regression coefficients, which was also verified by looking at Figure \ref{fig::real-estimates}. In addition,
comparing the estimated curve for the baseline intensities with the real curves of 10 observations, it is clear that the majority of the peaks and shapes are captured by the nonparametric estimate with appropriate degree of smoothing.

\begin{table}[t] \centering 
\caption{Regression parameters of real data} \label{para2}
\begin{center}
\begin{tabular}{c|r|r||c|r|r||c|r|r}
\hline
ID&$\hat\alpha $& $\hat\beta$&  ID&$\hat\alpha $& $\hat\beta$&  ID&$\hat\alpha $& $\hat\beta$\\
\hline
1&	0&	1&   12&	-1.0302&	1.5914& 23&	-0.7234&	1.4448\\
2&	-0.2086&	1.1836& 13&	-0.1788&	1.1366&    24&	0.2021&	0.8915\\
3&	-1.2208&	1.6836& 14&	-0.3252&	1.2586&    25&	0.5341&	0.7957\\
4&	-0.5630&	1.3689& 15&	-0.6169&	1.3599&    26&	0.53727&	0.7203\\
5&	-1.4761&	1.8721& 16&	-0.3919&	1.2418&    27&	-0.3748&	1.2181\\
6&	-1.2931&	1.7142& 17&	0.0820&	1.0178& 28&	0.0935&	0.9642\\
7&	0.7928&	0.5925&  18&	0.7402&	0.6569& 29&	0.7852&	0.5971\\
8&	0.0582&	1.0387&  19&	-0.0609&	1.0586&    30&	-0.0085&	 1.0503\\
9&	-0.3338&	1.1839& 20&	-0.9218&	1.5053&    31&	-0.3827&	1.2131\\
10&	-0.9066&	1.5397&    21&	-0.0580&	1.0378&    32&	-0.7803&	 1.5011\\
11&	-1.5054&	1.8770&    22&	-0.3526&	1.2149&    33&	-0.2115&	 1.1108\\

\hline
\end{tabular}
\end{center}
\end{table}

\section{Simulation Studies}\label{sec::simulation}

We conduct simulations to assess the performance of the proposed method for  parameter and curve estimation.
The true curve $m(\cdot)$ is chosen from a moving
average smoother of the cross-sectional mean of a fraction of real
Mass Spectrometry data in Section \ref{sec::realdata} after log transformation. We set 10000 $m/z$ values
equally-spaced from 1 to 10000 ($T=10000$) and the number of individuals  $n=30$. The true values of the parameters $\alpha_i,\beta_i, i = 1, 2, \ldots , n$ for each individual are shown in Table \ref{para1}. And the error terms $\epsilon_{it}$ are sampled independently from $N(0, \sigma^2)$ with $\sigma=0.25$.

\begin{figure}[t]
\caption{Nonparametric estimates of the curve from $\tilde m(\cdot)$ and $\hat m(\cdot)$ for the 16th object.\label{simu-estimate}}
    \begin{center}
      \includegraphics[scale=0.5]{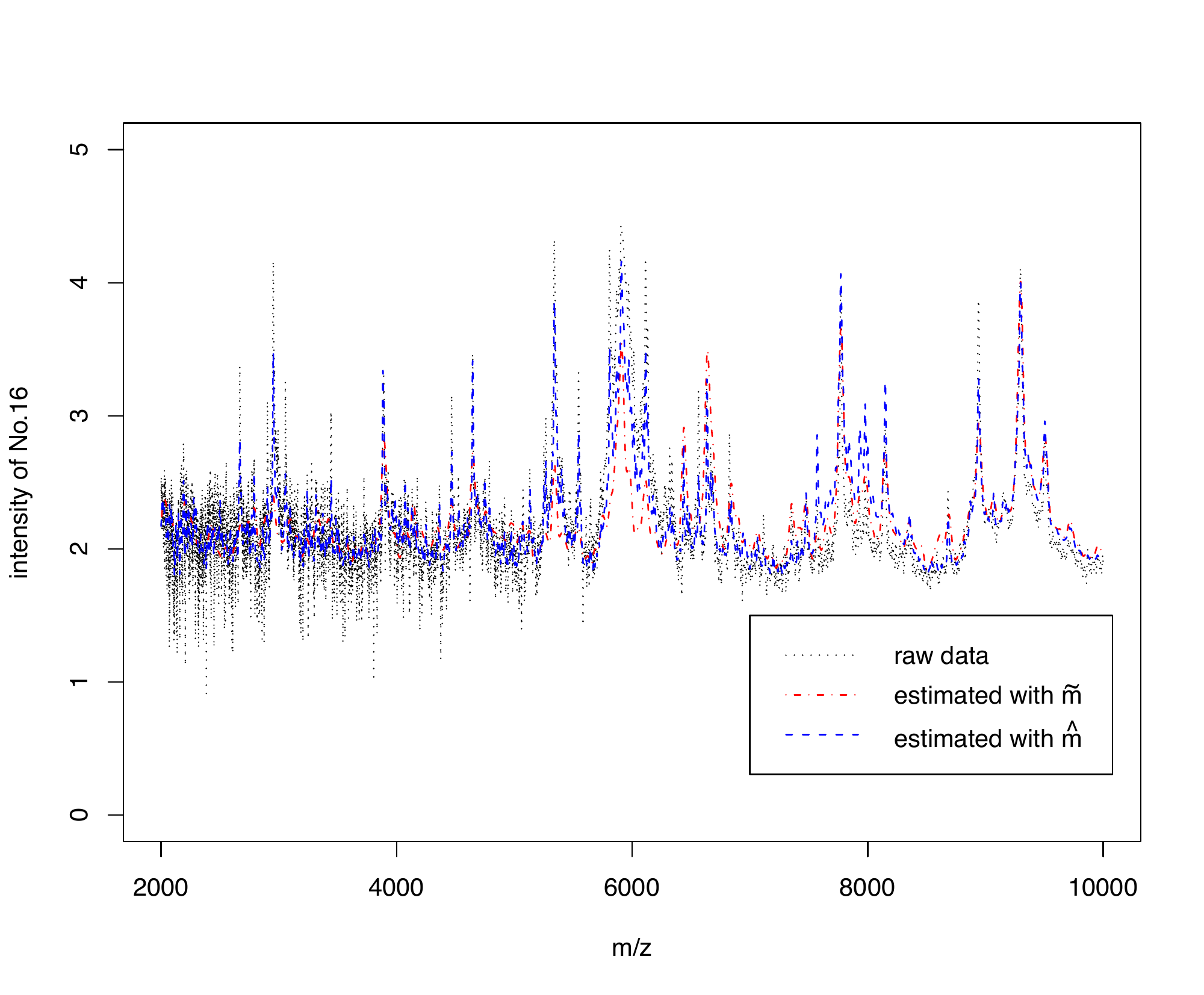}
    \end{center}
\vskip-8pt \centerline{\scriptsize m/z}
\end{figure}

We apply our multi-step procedure to the simulated data with different choices of the bandwidth. The number of runs is 100.  The estimated parameters $\hat\alpha_i$ and $\hat\beta_i$ are shown in Table \ref{para1} along with the standard errors. We set $h=35$, which leads to the smallest MSE of $\tilde m$ shown in Table \ref{tb::simu:mse}.  It is evident that the estimation is very accurate for all the location and scale parameters.  A graphical representation of the raw curve of the 16th subject along with estimates derived from $\tilde m(\cdot)$ and $\hat m(\cdot)$ can be found in Figure \ref{simu-estimate}. We can see from the figure that the estimate from $\hat m(\cdot)$ is notably better than that from $\tilde m(\cdot)$, which shows that multi-step procedure is effective in improving the estimates for the baseline curve. We observed similar phenomenon for all the other subjects.

From Table \ref{tb::simu:mse}, we can see that the global optimal bandwidths are $h=25,h^*=36$. It is interesting to note that the optimal bandwidth for $\tilde m(\cdot)$ is $h=35$, which is different from the optimal bandwidth for the final estimator.

To evaluate the quality of the our multi-step estimation method for the nonparametric baseline function, we consider a classical nonparametric estimation on another set of data where the same true function $m(\cdot)$ is used but $\alpha_i=0, \beta_i=1$ for all $i=1,\ldots,n$.  We applied the same local linear estimation with different bandwidths. The mean MSE of the estimated $m(\cdot)$ from 100 repetitions for different $h$s are given in Table \ref{mse2}. When we applied the multi-step estimation procedure, the best mean MSE we achieved in Table \ref{tb::simu:mse} is very close to the minimal mean MSE 0.4442 for the oracle estimator. This comparison confirms that there is little loss of information in the proposed method when both parametric and nonparametric components are estimated simultaneously.

\begin{table}[t] \centering
\caption{Regression parameter estimates ($h=35$)} \label{para1}
\begin{center}
\begin{tabular}{c|rr|c|c||c|rr|c|c}
\hline
ID&$\alpha$& $\beta$& $\hat\alpha$& $\hat\beta$&    ID&$\alpha$& $\beta$& $\hat\alpha$& $\hat\beta$\\
\hline
1&0&1&0.000(0.000)&1.000(0.000)&  16&0.6&1&0.605(0.026)&0.997(0.019)\\
2&0.2&0.2&0.202(0.020)&0.198(0.013)&    17&0.8&0.2&0.799(0.020)&0.201(0.014)\\
3&0.4&0.5&0.399(0.023)&0.501(0.016)&    18&1&0.5&1.004(0.024)&0.497(0.016)\\
4&0.6&1.5&0.598(0.040)&1.502(0.027)&    19&0&1.5&-0.001(0.044)&1.501(0.029)\\
5&0.8&2&0.801(0.047)&2.000(0.031)&  20&0.2&2&0.206(0.043)&1.997(0.029)\\
6&1&1&0.999(0.029)&1.001(0.020)&    21&0.4&1&0.403(0.030)&0.998(0.021)\\
7&0&0.2&-0.001(0.022)&0.201(0.015)& 22&0.6&0.2&0.598(0.024)&0.201(0.016)\\
8&0.2&0.5&0.200(0.021)&0.500(0.014)&    23&0.8&0.5&0.801(0.024)&0.500(0.016)\\
9&0.4&1.5&0.404(0.033)&1.498(0.023)&    24&1&1.5&0.996(0.038)&1.503(0.025)\\
10&0.6&2&0.603(0.044)&1.999(0.030)& 25&0&2&0.001(0.044)&2.000(0.029)\\
11&0.8&1&0.800(0.026)&1.001(0.018)& 26&0.2&1&0.203(0.030)&0.998(0.020)\\
12&1&0.2&1.002(0.021)&0.198(0.014)& 27&0.4&0.2&0.399(0.021)&0.201(0.015)\\
13&0&0.5&0.003(0.023)&0.499(0.016)& 28&0.6&0.5&0.604(0.021)&0.497(0.014)\\
14&0.2&1.5&0.206(0.036)&1.497(0.024)&   29&0.8&1.5&0.803(0.032)&1.499(0.023)\\
15&0.4&2&0.401(0.048)&2.001(0.033)& 30&1&2&1.001(0.047)&2.000(0.031)\\

\hline \multicolumn{7}{l}{$^*$Standard deviations are in
parentheses}
\end{tabular}
\end{center}
\end{table}

\begin{table}[t] \centering 
\caption{MSE of the initial and updated estimation of $m$} \label{tb::simu:mse}
\begin{center}
\begin{tabular}{c|r|r|r|r|r}
\hline
\multicolumn{6}{c}{MSE of $\tilde m$}\\
\hline
$h$&20&25&30&35&40\\
\hline
 &9.1112&7.5509&6.7024&\textbf{6.3418}&6.3653\\
\hline
\hline
\multicolumn{6}{c}{MSE of $\tilde m$}\\
\hline
\backslashbox{$h^*$}{$h$} &20&25&30&35&40\\
\hline
20&0.6145&0.5936&0.5925&0.6078&0.6388\\
22&0.5762&0.5563&0.5561&0.5723&0.6042\\
24&0.5453&0.5265&0.5272&0.5443&0.5771\\
26&0.5204&0.5026&0.5043&0.5223&0.5561\\
28&0.5005&0.4838&0.4866&0.5056&0.5403\\
30&0.4850&0.4695&0.4733&0.4934&0.5291\\
32&0.4735&0.4592&0.4641&0.4852&0.5220\\
34&0.4657&0.4527&0.4587&0.4809&0.5188\\
36&0.4612&\textbf{0.4496}&0.4568&0.4802&0.5192\\
38&0.4601&0.4498&0.4583&0.4829&0.5231\\
40&0.4622&0.4533&0.4631&0.4889&0.5303\\
\hline
\end{tabular}
\end{center}
\end{table}

\begin{table}[t] \centering 
\caption{MSE of the estimation of $m$ in the dataset with same parameters in each individual} \label{mse2}
\begin{center}
\begin{tabular}{c|r|r|r|r|r}
\hline
h&20&	30&	40&	50&	60\\
\hline
MSE&0.6881&	0.4936&	\textbf{0.4442}&	0.4926&	0.6337\\
\hline
\end{tabular}
\end{center}
\end{table}


We use cross-validation to get a data-driven choice of the bandwidths. Here,  we set $K=5$  to get a mean MSPE of every different bandwidth choices of both steps over 100 runs, and the optimal bandwidths are those with the minimum mean MSPE. The mean MSPE values are shown in  Table \ref{tb::simu.mspe.cv}, from which we can see that the smallest value is located at $h=25,h^*=36$, which is quite close to the optimal bandwidths $h=25$ and $h^*=38$ in Table \ref{tb::simu:mse}. Therefore, the cross-validation idea appears to work well in terms of selecting the best bandwidths.

\begin{table}[t] \centering 
\caption{Mean MSPE of the 5-fold CV over 100 times. Here, we multiply $T$ for all the MSPE  values and subtract the minimum 625.61956.} \label{tb::simu.mspe.cv}
\begin{center}
\begin{tabular}{c|r|r|r|r|r}
\hline
\backslashbox{$h^*$}{$h$}&20&25&30&35&40\\
\hline
20&	0.293733&	0.293728&	0.293724&	0.293722&	0.293721\\
22&	0.222948&	0.222943&	0.222940&	0.222939&	0.222939\\
24&	0.165816&	0.165813&	0.165810&	0.165809&	0.165809\\
26&	0.119253&	0.119250&	0.119248&	0.119248&	0.119248\\
28&	0.081603&	0.081600&	0.081599&	0.081598&	0.081599\\
30&	0.052297&	0.052295&	0.052294&	0.052294&	0.052295\\
32&	0.030256&	0.030255&	0.030254&	0.030255&	0.030256\\
34&	0.014621&	0.014620&	0.014620&	0.014621&	0.014622\\
36&	0.004761&	0.004760&	0.004760&	0.004761&	0.004763\\
38&	$3.4e-08$&	$0$&	$5.6e-07$&	$2.0e-06$&	$4.2e-06$\\
40&	0.000590&	0.000590&	0.000592&	0.000593&	0.000596\\

\hline
\end{tabular}
\end{center}
\end{table}

\section{Discussion}
This paper proposes a semi- and nonparametric model suitable for analyzing the mass spectra data. The model is flexible and intuitive, capturing the main feature in the MS data. Both the parametric and nonparametric components have natural interpretation.  A multi-step iterative algorithm is proposed for estimating both the individual location and scale regression coefficients and the nonparametric function. The algorithm combines local linear fitting and the least squares method, both of which are easy to implement and computationally efficient.   Both simulation studies and real data analysis demonstrate that the proposed multi-step procedure works well.

The local linear fitting for the nonparametric function estimation maybe replaced with other nonparametric estimation techniques. Because the location and scale parameters are subject specific, the empirical Bayes method \citep{Carlon-Louis-2008} may be used. In addition, nonparametric Bayes may also be applicable with the nonparametric function being modeling as a realization of Gaussian process.

 The proposed model and the associated iterative estimation method do not account for the random error in the measurement of $X$. It is desirable to incorporate the measurement error into the model \citep{Carroll-2006} .

 Many studies involving MS data are aimed at classifying patients of different disease types. The information of peaks are usually applied as the basis of the classifier. The proposed method provides a natural way of finding the peaks for different group of patients by use the multi-step estimation procedure on each group and find out the corresponding nonparametric baseline function. From the estimated baseline function, the information of peaks can be easily extracted, which can then be used for classification.

\section*{Acknowledgement}
The research was supported in part by National Institutes of Health grant R37GM047845. The authors would like to thank Liang Zhu and Cheng Wu at Shanghai Changzheng Hospital for providing
the data.

\section*{Appendix}
The Appendix contains proofs of Lemma 1, Corollary 1 and Theorems 1-4. We begin with
some notation, which will be used to streamline some of the proofs. Because all asymptotic expansions are derived with $x_{it}$'s being fixed, we will, for notational simplicity, use $E$ to denote the conditional expectation and $\Var$ to denote the conditional variance given $x_{it}$'s throughout the Appendix. For $i=1,\ldots, n$ and $t=1,\ldots,T$, let

$$V_{it}=\frac{1}{T}-\frac{\sigma_i(X_{it})W_{it}\bar m(x_{i\cdot})}{\sum_{s=1}^TW_{is}^2}.$$


\subsection{Proof of Lemma \ref{lm1}}

\begin{proof}
It follows from \eqref{eq::S_Tk} and the definition of $W_{it}$ that  $$\sum_{t=1}^T\omega_{1t}(x)(x_{1t}-x)=S_{T,2}S_{T,1}-S_{T,1}S_{T,2}=0.$$
From Condition C1, we have
\begin{eqnarray*}
m(x_{1t})=m(x)+m'(x)(x_{1t}-x)+\frac{1}{2}m''(x)(x_{1t}-x)^2+o((x_{1t}-x)^2),
\end{eqnarray*}
where $o(\cdot)$ is uniform in $t$.
Thus
\begin{align}
\tilde m(x)
&=\frac{\sum_{t=1}^T\omega_{1t}(x)[m(x)+\frac{1}{2}m''(x)(x_{1t}-x)^2+o((x_{1t}-x)^2)]}{\sum_{t=1}^T\omega_{1t}(x)}\nonumber\\
&\quad +\frac{\sum_{t=1}^T\omega_{1t}(x)\sigma_1(x_{1t})\epsilon_{1t}}{\sum_{t=1}^T\omega_{1t}(x)}\nonumber\\
&=m(x)+\frac{\sum_{t=1}^T\omega_{1t}(x)[\frac{1}{2}m''(x)(x_{1t}-x)^2+o((x_{1t}-x)^2)]}{\sum_{t=1}^T\omega_{1t}(x)}+U_1(x)\nonumber\\
&=m(x)+\frac{(S_{T,2}^2-S_{T,1}S_{T,3})m''(x)}{2(S_{T,0}S_{T,2}-S_{T,1}^2)}+o\Big(\frac{S_{T,2}^2-S_{T,1}S_{T,3}}{S_{T,0}S_{T,2}-S_{T,1}^2}\Big)+U_1(x),\label{eq::tildem}
\end{align}
where the last equality follows from the definition of $S_{T,j}^2, j=0,1,2,3$.


A standard asymptotic expansion for the local linear smoothing (Fan and Gijbels, 1996, eq 3.13) results in
\begin{equation}\label{eq::Stj:expansion}
  S_{T,j}=Th^jf(x)\mu_j\{1+o_P(1)\}, j=0,1,2,3.
\end{equation}
Note that with $j=0$ and 1 in \eqref{eq::Stj:expansion}, we have, $S_{T,0}=Tf(x)(1+o_p(1))$ and $S_{T,1}=o_p(1)$ since $\mu_0=1$ and $\mu_1=0$, combined with \eqref{eq::tildem},  \begin{eqnarray*}
\tilde m(x)=m(x)+\frac{1}{2}m''(x)\mu_2h^2+o(h^2)+U_1(x).
\end{eqnarray*}
This completes the proof of  Lemma \ref{lm1}.
\end{proof}

\subsection{Proof of Corollary \ref{co1}}
\begin{proof}
Being a weighted average of mean-zero random variables, $U_1(x)$ has zero mean.
Thus, from Lemma 1,  we have $$\E[\tilde m(x)-m(x)]=\frac{1}{2}m''(x)\mu_2h^2+o(h^2).$$
For the variance term, from the definition of $\tilde m(\cdot)$,  we have
\small
\begin{align*}
\Var(\tilde m(x))&=\Var\left(\frac{\sum_{t=1}^T\omega_{1t}(x)\sigma_1(x_{1t})\epsilon_{1t}}{\sum_{t=1}^T\omega_{1t}(x)}\right)\\
&=\Var\left(\frac{S_{T,2}\sum_{t=1}^TK_h(x_{1t}-x)\sigma_1(x_{1t})\epsilon_{1t}-S_{T,1}\sum_{t=1}^TK_h(x_{1t}-x)(x_{1t}-x)\sigma_1(x_{1t})\epsilon_{1t}} {S_{T,0}S_{T,2}-S_{T,1}^2}\right)\\
&=\Var\left(\frac{\sum_{t=1}^TK_h(x_{1t}-x)\sigma_1(x_{1t})\epsilon_{1t}}{S_{T,0}}\right)+o\left(\Var\biggm(\frac{\sum_{t=1}^TK_h(x_{1t}-x)\sigma_1(x_{1t})\epsilon_{1t}}{S_{T,0}}\biggm)\right)\\
&=\frac{1}{Th}[f(x)]^{-1}\sigma_1^2(x)\nu_0+o\Big(\frac{1}{Th}\Big),
\end{align*}
where the third equation follows from \eqref{eq::Stj:expansion}, and the last equation follows from Condition C3 and \eqref{eq::Stj:expansion}.
Combining the above asymptotic expansions for the bias and variance terms leads to the desired expansion for  the mean squared error.
\end{proof}


\subsection{Proof of Theorem 1}

\begin{proof}
First of all, define $\tilde W_{it}=\tilde m(x_{it})-\bar{\tilde{ m}}(x_{i\cdot})$ to simplify the presentation.
By definition, we have the following expansion for $\hat\beta_i$ when $i\geq 2$.
\begin{eqnarray}
\hat\beta_i-\beta_i
&=&\beta_i\frac{\sum_{t=1}^T\tilde W_{it}(m(x_{it})-\tilde m(x_{it}))}{\sum_{t=1}^T\tilde W_{it}^2} +\frac{\sum_{t=1}^T\tilde W_{it}\sigma_i(x_{it})\epsilon_{it}}{\sum_{t=1}^T\tilde W_{it}^2}\nonumber\\
&\equiv&\beta_iD_i+G_i.\label{eq::beta_i-bias}
\end{eqnarray}
From Lemma 1 and the proof of Corollary \ref{co1}, we have
\begin{align}\label{eq:tildeW-W}
\tilde W_{it}-W_{it}=O_p(h^2).
\end{align}

Plugging (\ref{eq-lm1}) into $D_i$, we have

\begin{align}
D_i
&=-\frac{\sum_{t=1}^T[(U_1(x_{it})-\bar U_1(x_{i\cdot}))U_1(x_{it})+\frac{1}{2}\mu_2m''(x_{it})W_{it}h^2+o(h^2)]    }{\sum_{t=1}^T \tilde W_{it} ^2}\notag\\
&\quad-\frac{\sum_{t=1}^T\{[W_{it}+O(h^2)]U_1(x_{it})+O(h^2)[U_1(x_{it})-\bar U_1(x_{i\cdot})]\}}{\sum_{t=1}^T \tilde W_{it} ^2}\notag\\
&=-\frac{\sum_{t=1}^T(U_1(x_{it})-\bar U_1(x_{i\cdot}))U_1(x_{it})}{\sum_{t=1}^TW_{it}^2}-h^2P_i-\frac{\sum_{t=1}^TW_{it}U_1(x_{it})}{\sum_{t=1}^TW_{it}^2}(1+o_p(1))\notag\\
&\quad+o(h^2+\frac{1}{Th}),\label{eq::D_i-expansion}
\end{align}
where the last asymptotic expansion follows from \eqref{eq:tildeW-W}.
Similarly for $G_i$, we have

\begin{align}
G_i&=\frac{\sum_{t=1}^TW_{it}\sigma_i(x_{it})\epsilon_{it}(1+O(h^2))}{\sum_{t=1}^T\tilde W_{it}^2}+\frac{\sum_{t=1}^T(U_1(x_{it})-\bar U_1(x_{i\cdot}))\sigma_i(x_{it})\epsilon_{it}}{\sum_{t=1}^T\tilde W_{it}^2}\notag\\
&=\frac{\sum_{t=1}^TW_{it}\sigma_i(x_{it})\epsilon_{it}}{\sum_{t=1}^TW_{it}^2}(1+o_p(1)).\label{eq::G_i-expansion}
\end{align}



We observe that for any $i\geq 2$, $U_1(x_{it})$  is a  linear combination of $\{\epsilon_{1t}, t=1,\ldots, T\}$. Therefore, $U_1(x_{it})$ is  independent of $\{\epsilon_{it}, i=2,\ldots,n, t=1,\ldots,T\}$. By using the tower property, we have $EG_i= 0$. Therefore, $\beta_iD_i$ is the only part that contributes to the bias of $\hat\beta_i$. In view of these and Corollary \ref{co1}, we have the following expansions for the bias and variance terms
\begin{align*}
\E(\hat\beta_i-\beta_i)
&=-\beta_ih^2P_i-\beta_i\frac{\E\sum_{t=1}^T(U_1(x_{it})-\bar U_1(x_{i\cdot}))^2}{\sum_{t=1}^TW_{it}^2}+o(h^2+\frac{1}{Th})\\
&=-\beta_ih^2P_i-\beta_i\frac{\sum_{t=1}^T\Var(U_1(x_{it}))(1+o(1))}{\sum_{t=1}^TW_{it}^2}+o(h^2+\frac{1}{Th})\\
&=-\beta_ih^2P_i-\beta_i\frac{1}{Th}Q_i+o(h^2+\frac{1}{Th}),
\end{align*}
and
\begin{align*}
\Var(\hat\beta_i)=& \Var\left(-\beta_i\frac{\sum_{t=1}^TW_{it}U_1(x_{it})}{\sum_{t=1}^TW_{it}^2}(1+o_p(1))\right)\\
+& \Var\left(\frac{\sum_{t=1}^TW_{it}\sigma_i(x_{it})\epsilon_{it}}{\sum_{t=1}^TW_{it}^2}(1+o_p(1))\right).
\end{align*}
\normalsize
%


Straightforward variance calculation for an independent sum gives

\begin{align}\label{eq:varwit}
\Var\left(\sum_{t=1}^TW_{it}U_1(x_{it})\right)
&=\sum_{s=1}^T\Bigm[\sum_{t=1}^TW_{it}\frac{\omega_{1s}(x_{it})}{\sum_{l=1}^T\omega_{1l}(x_{it})}\Bigm]^2\sigma_1^2(x_{1s}).
\end{align}


We have

\begin{align*}
&\sum_{t=1}^TW_{it}\frac{\omega_{1s}(x_{it})}{\sum_{s=1}^T\omega_{1s}(x_{it})}\\
=&\sum_{t=1}^T\big(m(x_{it})-\bar m(x_{i\cdot})\big)\frac{K_h(x_{it}-x_{1s})(S_{T,2}-(x_{it}-x_{1s})S_{T,1})}{S_{T,0}S_{T,2}-S_{T,1}^2} .\\
\end{align*}

We expand $m(x)$ in the neighborhood of point $x_{1s}$ using Taylor's expansion, $$m(x_{it})=m(x_{1s})+(x_{it}-x_{1s})m'(x_{1s})+\frac{1}{2}(x_{it}-x_{1s})^2m''(x_{1s})+o_p((x_{it}-x_{1s})^2).$$
Since the kernel function $K_h(x-x_{1s})$ vanishes out of the neighborhood of $x_{1s}$ with diameter $h$, we can obtain the following
\begin{align*}
&\sum_{t=1}^Tm(x_{it})\frac{K_h(x_{it}-x_{1s})(S_{T,2}-(x_{it}-x_{1s})S_{T,1})}{S_{T,0}S_{T,2}-S_{T,1}^2}\\
=& m(x_{1s})+\sum_{t=1}^Tm'(x_{1s})(x_{it}-x_{1s})\frac{K_h(x_{it}-x_{1s})(S_{T,2}-(x_{it}-x_{1s})S_{T,1})}{S_{T,0}S_{T,2}-S_{T,1}^2}\\
&+\sum_{t=1}^T\big[\frac{1}{2}(x_{it}-x_{1s})^2m''(x_{1s})+o_p((x_{it}-x_{1s})^2)\big]\frac{K_h(x_{it}-x_{1s})(S_{T,2}-(x_{it}-x_{1s})S_{T,1})}{S_{T,0}S_{T,2}-S_{T,1}^2}\\
=& m(x_{1s})+O_p(h^2)\sum_{t=1}^T\frac{K_h(x_{it}-x_{1s})(S_{T,2}-(x_{it}-x_{1s})S_{T,1})}{S_{T,0}S_{T,2}-S_{T,1}^2}\\
=& m(x_{1s})+O_p(h^2),
\end{align*}
where the functions $S_{T,k}, k=0,1,2$ are evaluated at the point $x_{it}$.
Combined with $\bar m(x_{i\cdot})=\bar m(x_{1\cdot})+O_p(T^{-1/2})$, we can have the expansion $$\sum_{t=1}^TW_{it}\frac{\omega_{1s}(x_{it})}{\sum_{s=1}^T\omega_{1s}(x_{it})} = m(x_{1s})+O_p(h^2)-\bar m(x_{1\cdot})+O_p(T^{-1/2}) = W_{1s}+O_p(h^2+T^{-1/2})$$

%

Then recall \eqref{eq:varwit}, we have $\Var(\sum_{t=1}^TW_{it}U_1(x_{it}))=\sum_{t=1}^TW_{1t}^2\sigma_1^2(x_{1t})+o_p(T)$,
which leads to the variance expansion

\begin{align*}
\Var(\hat\beta_i)=\beta_i^2\frac{\sum_{t=1}^TW_{1t}^2\sigma_1^2(x_{1t})}{(\sum_{t=1}^TW_{it}^2)^2}+\frac{\sum_{t=1}^TW_{it}^2\sigma_i^2(x_{it})}{(\sum_{t=1}^TW_{it}^2)^2}+o_p\Big(\frac{1}{T}\Big).
\end{align*}

\end{proof}

\subsection{Proof of Theorem 2}
\begin{proof}
Recall \eqref{eq::omegastar} and  \eqref{eq::S_Tkstar}, we have $$\sum_{i=1}^n\sum_{t=1}^T\omega_{it}^*(x)(x_{it}-x)=\sum_{i=1}^n\hat\beta_i^2S_{T,2}^{*(i)}\sum_{i=1}^n\hat\beta_i^2S_{T,1}^{*(i)}-\sum_{i=1}^n\hat\beta_i^2S_{T,1}^{*(i)}\sum_{i=1}^n\hat\beta_i^2S_{T,2}^{*(i)}=0.$$

Then we have the asymptotic expansion of the updated estimator of baseline intensity $\hat m(\cdot)$ at time point $x$ as follows.

By definition of $\hat m(\cdot)$ in \eqref{eq::hat_m} , we can write
\begin{align}
&\hat m(x)-m(x)\nonumber\\
=&\frac{\sum_{i=1}^n\sum_{t=1}^T\omega_{it}^* (\alpha_i-\hat\alpha_i)\big/\hat\beta_i }{\sum_{i=1}^n\sum_{t=1}^T\omega_{it}^*}+
\frac{\sum_{i=1}^n\sum_{t=1}^T\omega_{it}^* \sigma(x_{it})\epsilon_i \big/\hat\beta_i}{\sum_{i=1}^n\sum_{t=1}^T\omega_{it}^*}\nonumber\\
&\qquad+\frac{\sum_{i=1}^n\sum_{t=1}^T\omega_{it}^* m(x_{it})(\beta_i-\hat\beta_i)\big/\hat\beta_i }{\sum_{i=1}^n\sum_{t=1}^T\omega_{it}^*}\nonumber\\
&\qquad+\frac{\sum_{i=1}^n\sum_{t=1}^T\omega_{it}^* (\frac{1}{2}m''(x)(x_{it}-x)^2+o((x_{it}-x)^2) ) }{\sum_{i=1}^n\sum_{t=1}^T\omega_{it}^*}.\label{eq::hat_m_bias}
\end{align}


From the proof of  Theorem \ref{th1}, we have
\small
\begin{align}
\hat\beta_i
&=\beta_i-\beta_ih^2P_i-\beta_i\frac{\sum_{t=1}^T(U_1(x_{it})-\bar U_1(x_{i\cdot}))U_1(x_{it})}{\sum_{t=1}^TW_{it}^2}-\beta_i\frac{\sum_{t=1}^TW_{it}U_1(x_{it})}{\sum_{t=1}^TW_{it}^2}(1+o_p(1))\nonumber\\
&\quad+\frac{\sum_{t=1}^TW_{it}\sigma_i(x_{it})\epsilon_{it}}{\sum_{t=1}^TW_{it}^2}(1+o_p(1))+o\Big(h^2+\frac{1}{Th}\Big).\label{eq::hat_beta_expansion}
\end{align}
Then, from the least square expression, we have the asymptotic expansion for $\hat \alpha_i$ as follows.
\begin{align}
\hat\alpha_i
&=\bar y_{i\cdot}-\hat\beta_i\bar{\tilde m}(x_{i\cdot})\nonumber\\
&=\alpha_i+\beta_i\bar m(x_{i\cdot})+\bar\epsilon_{i\cdot}-\hat\beta_i[\bar m(x_{i\cdot})+\frac{\mu_2}{2}\bar{ m''}(x_{i\cdot})h^2+\bar U_1(x_{i\cdot})+o(h^2)]\nonumber\\
&=\alpha_i+\beta_ih^2R_i+\bar m(x_{i\cdot})\beta_i\frac{\sum_{t=1}^T(U_1(x_{it})-\bar U_1(x_{i\cdot}))U_1(x_{it})}{\sum_{t=1}^TW_{it}^2}+o\Big(h^2+\frac{1}{Th}\Big)\nonumber\\
&\quad+\sum_{t=1}^TV_{it}\epsilon_{it}(1+o_p(1))-\beta_i\sum_{t=1}^TV_{it}U_1(x_{it})(1+o_p(1)).\label{eq::hat_alpha_expansion}
\end{align}

Now, we plug the above asymptotic expansions \eqref{eq::hat_beta_expansion} and \eqref{eq::hat_alpha_expansion} into the right hand side of \eqref{eq::hat_m_bias}.
%
The first part of \eqref{eq::hat_m_bias} could be expanded as follows
\begin{align}
&\frac{\sum_{i=1}^n\sum_{t=1}^T\omega_{it}^* (\alpha_i-\hat\alpha_i)\big/\hat\beta_i }{\sum_{i=1}^n\sum_{t=1}^T\omega_{it}^*}\nonumber\\
=&\frac{\sum_{i=1}^n\sum_{t=1}^T(\alpha_i-\hat\alpha_i)\big/\hat\beta_i*\hat\beta_i^2K_{h^*}(x_{it}-x) \left[\sum_{i=1}^n\hat\beta_i^2S_{T,2}^{*(i)}-(x_{it}-x)\sum_{i=1}^n\hat\beta_i^2S_{T,1}^{*(i)}\right]} {\sum_{i=1}^n\sum_{t=1}^T\hat\beta_i^2K_{h^*}(x_{it}-x)\left[\sum_{i=1}^n\hat\beta_i^2S_{T,2}^{*(i)}-(x_{it}-x)\sum_{i=1}^n\hat\beta_i^2S_{T,1}^{*(i)}\right]}\nonumber\\
=&\frac{\sum_{i=1}^n(\alpha_i-\hat\alpha_i)\hat\beta_i\sum_{t=1}^TK_{h^*}(x_{it}-x) \left[\sum_{i=1}^n\hat\beta_i^2S_{T,2}^{*(i)}-(x_{it}-x)\sum_{i=1}^n\hat\beta_i^2S_{T,1}^{*(i)}\right]} {\sum_{i=1}^n\hat\beta_i^2\sum_{t=1}^TK_{h^*}(x_{it}-x)\left[\sum_{i=1}^n\hat\beta_i^2S_{T,2}^{*(i)}-(x_{it}-x)\sum_{i=1}^n\hat\beta_i^2S_{T,1}^{*(i)}\right]}.\label{eq::hat_m_bias part 1}
\end{align}

The numerator of \eqref{eq::hat_m_bias part 1} has expansion
\begin{align}
&\sum_{i=1}^n(\alpha_i-\hat\alpha_i)\hat\beta_i\bigg[Tf(x)\{1+o_p(1)\} \sum_{i=1}^n\hat\beta_i^2Th^{*2}f(x)\mu_2\{1+o_p(1)\}\nonumber\\
&-Th^*\{h^*f'(x)\mu_2+O_p(h^{*2}+\frac{1}{\sqrt{Th^*}})\}\sum_{i=1}^n\hat\beta_i^2Th^*\{h^*f'(x)\mu_2+O_p(h^{*2}+\frac{1}{\sqrt{Th^*}})\}\bigg]\nonumber\\
=&T^2h^{*2}\sum_{i=1}^n(\alpha_i-\hat\alpha_i)\hat\beta_i\bigg[f(x)\{1+o_p(1)\} \sum_{i=1}^n\hat\beta_i^2f(x)\mu_2\{1+o_p(1)\}\nonumber\\
&-\{h^*f'(x)\mu_2+O_p(h^{*2}+\frac{1}{\sqrt{Th^*}})\}\sum_{i=1}^n\hat\beta_i^2\{h^*f'(x)\mu_2+O_p(h^{*2}+\frac{1}{\sqrt{Th^*}})\}\bigg]\nonumber\\
=&T^2h^{*2}\sum_{i=1}^n(\alpha_i-\hat\alpha_i)\hat\beta_i\bigg[f(x)\sum_{i=1}^n\beta_i^2f(x)\mu_2\{1+o_p(1)\}\bigg],\label{eq:hat_m_bias_p1_nume}
\end{align}
where the last equation following from $\hat\beta_i=\beta_i+O(h^2)+O((Th)^{-1})+O_p(T^{-1/2})$.

Similarly, the denominator of \eqref{eq::hat_m_bias part 1} has the following expansion
\begin{align}
&\sum_{i=1}^n\hat\beta_i^2\bigg[Tf(x)\{1+o_p(1)\} \sum_{i=1}^n\hat\beta_i^2Th^{*2}f(x)\mu_2\{1+o_p(1)\}\nonumber\\
&-Th^*\{h^*f'(x)\mu_2+O_p(h^{*2}+\frac{1}{\sqrt{Th^*}})\}\sum_{i=1}^n\hat\beta_i^2Th^*\{h^*f'(x)\mu_2+O_p(h^{*2}+\frac{1}{\sqrt{Th^*}})\}\bigg]\nonumber\\
=&T^2h^{*2}\sum_{i=1}^n\hat\beta_i^2\bigg[f(x)\{1+o_p(1)\} \sum_{i=1}^n\hat\beta_i^2f(x)\mu_2\{1+o_p(1)\}\nonumber\\
&-\{h^*f'(x)\mu_2+O_p(h^{*2}+\frac{1}{\sqrt{Th^*}})\}\sum_{i=1}^n\hat\beta_i^2\{h^*f'(x_i)\mu_2+O_p(h^{*2}+\frac{1}{\sqrt{Th^*}})\}\bigg]\nonumber\\
=&T^2h^{*2}\sum_{i=1}^n\hat\beta_i^2\bigg[f(x)\sum_{i=1}^n\beta_i^2f(x)\mu_2\{1+o_p(1)\}\bigg].\label{eq:hat_m_bias_p1_deno}
\end{align}

Then combining the expansions \eqref{eq:hat_m_bias_p1_nume} and \eqref{eq:hat_m_bias_p1_deno}, we have the following expansion for the first part of \eqref{eq::hat_m_bias}.
\begin{align*}
&\frac{\sum_{i=1}^n\sum_{t=1}^T\omega_{it}^* (\alpha_i-\hat\alpha_i)\big/\hat\beta_i }{\sum_{i=1}^n\sum_{t=1}^T\omega_{it}^*}\nonumber\\ =&\frac{T^2h^{*2}\sum_{i=1}^n(\alpha_i-\hat\alpha_i)\hat\beta_i\bigg[f(x)\sum_{i=1}^n\beta_i^2f(x)\mu_2\{1+o_p(1)\}\bigg]}{T^2h^{*2}\sum_{i=1}^n\hat\beta_i^2\bigg[f(x)\sum_{i=1}^n\beta_i^2f(x)\mu_2\{1+o_p(1)\}\bigg]}\\
=&\frac{\sum_{i=2}^n(\alpha_i-\hat\alpha_i)\beta_i}{\sum_{i=1}^n\beta_i^2}(1+o_p(1)).
\end{align*}
For other parts of \eqref{eq::hat_m_bias}, we can apply the same techniques for expansion. As a result, the following expansion of $\hat m$ holds. 

%
\begin{align*}
&\hat m(x)-m(x)\\
&\quad=\frac{\sum_{i=2}^n\beta_i(\alpha_i-\hat\alpha_i)}{\sum_{i=1}^n\beta_i^2}(1+o_p(1))
+\frac{\sum_{i=1}^n\beta_i\sum_{t=1}^TK_{h^*}(x_{it}-x)\epsilon_{it}(1+o_p(1))}{\sum_{i=1}^n\beta_i^2Tf(x)}\\
&\qquad+\frac{\sum_{i=1}^n\beta_i(\beta_i-\hat\beta_i)\sum_{t=1}^TK_{h^*}(x_{it}-x)\epsilon_{it}(1+o_p(1))}{\sum_{i=1}^n\beta_i^2Tf(x)}\\
&\quad=-\frac{\sum\limits_{i=2}^n\beta_i^2\Bigm[   h^2R_i+\bar m(x_{i\cdot})\sum\limits_{t=1}^T(U_1(x_{it})-\bar U_1(x_{i\cdot}))^2/\sum\limits_{t=1}^TW_{it}^2-\sum\limits_{t=1}^TV_{it}U_1(x_{it})    (1+o_p(1))\Bigm]}{\sum_{i=1}^n\beta_i^2}\\
&\qquad+\frac{\sum_{i=1}^n\beta_i\sum_{t=1}^TK_{h^*}(x_{it}-x)\epsilon_{it}(1+o_p(1))}{\sum_{i=1}^n\beta_i^2}\\
&\qquad+m(x)\left\{\frac{\sum_{i=2}^n\beta_i^2h^2P_i}{\sum_{i=1}^n\beta_i^2}+\frac{\sum_{i=2}^n\beta_i^2\sum_{t=1}^T(U_1(x_{it})-\bar U_1(x_{i\cdot}))^2/\sum_{t=1}^TW_{it}^2}{\sum_{i=1}^n\beta_i^2f(x)}\right\}\\
&\qquad+m(x)\left\{\frac{\sum_{i=2}^n\beta_i^2\sum_{t=1}^TW_{it}U_1(x_{it})(1+o_p(1))/\sum_{t=1}^TW_{it}^2}{\sum_{i=1}^n\beta_i^2}\right\}\\
&\qquad+\frac{m''(x)}{2}\mu_2h^{*2}+o\Big(h^2+\frac{1}{Th}+h^{*2}\Big).
\end{align*}
\normalsize

Then it is straightforward to derive the bias of $\hat m(x)$ as follows
\small
\begin{align*}
&\E(\hat m(x)-m(x))\\
&\quad=  -\frac{\sum_{i=2}^n\beta_i^2(   h^2R_i+(Th)^{-1}\bar m(x_{i\cdot})Q_i )}{\sum_{i=1}^n\beta_i^2}+  \Bigm[\frac{\sum_{i=2}^n\beta_i^2(   h^2P_i+(Th)^{-1}Q_i )}{\sum_{i=1}^n\beta_i^2}\Bigm]  m(x)\\
&\qquad+\frac{m''(x)}{2}\mu_2h^{*2}+o\Big(h^2+\frac{1}{Th}+h^{*2}\Big).
\end{align*}
\normalsize
For the variance of $\hat m(x)$, we notice that the error terms $\{\epsilon_{it}, i=1,\ldots,n, t=1,\ldots, T\}$  are independent, which implies the independence of $\epsilon_{it},i=2,\ldots,n$ and $U_1(x_{it})$.  Therefore, we have the following asymptotic expansion for the variance.

\begin{align*}
&\Var(\hat m(x)-m(x)) \\
&\quad= \Var\left( \frac{\sum_{i=2}^n\beta_i^2\sum_{t=1}^TV_{it}U_1(x_{it}) }{\sum_{i=1}^n\beta_i^2}  +
\frac{m(x)\sum_{i=2}^n\beta_i^2\sum_{t=1}^TW_{it}U_1(x_{it})\big/\sum_{t=1}^TW_{it}^2}{\sum_{i=1}^n\beta_i^2} \right.\\
&\qquad\left.+  \frac{\sum_{i=1}^n\sum_{t=1}^T\beta_iK_{h^*}(x_{it}-x)\epsilon_{it}}{\sum_{i=1}^n\beta_i^2\sum_{t=1}^TK_{h^*}(x_{it}-x)}  \right)+o\Big(\frac{1}{T}+\frac{1}{nTh^*}\Big)\\
&\quad=\Var\left(\frac{\sum_{i=2}^n\beta_i^2\sum_{t=1}^T(V_{it}+m(x)W_{it}/\sum_{t=1}^TW_{it}^2)U_1(x_{it})}{\sum_{i=1}^n\beta_i^2}\right)\\
&\qquad+\frac{\nu_0\sum_{i=2}^n\beta_i^2f^{-1}(x)\sigma_i^2(x_{it})}{Th^*(\sum_{i=1}^n\beta_i^2)^2}+o\Big(\frac{1}{T}+\frac{1}{nTh^*}\Big),
\end{align*}
where the expansions follow similar techniques as \eqref{eq:hat_m_bias_p1_nume} and \eqref{eq:hat_m_bias_p1_deno}.
Now, by the definition of $U_1$, we have
\small
\begin{align*}
&\Var(\hat m(x)-m(x)) \\
&\quad=\frac{1}{(\sum_{i=1}^n\beta_i^2)^2}\sum_{s=1}^T\Bigm(\sum_{i=2}^n\beta_i^2\bigm[\frac{1}{T}+Z_{is}\bigm]\Bigm)^2\sigma_1^2(x_{1s})+\frac{\nu_0\sum_{i=2}^n\beta_i^2f^{-1}(x)\sigma_i^2(x_{it})}{Th^*(\sum_{i=1}^n\beta_i^2)^2}\\
&\qquad+o\Bigm(\frac{1}{T}+\frac{1}{nTh^*}\Bigm).
\end{align*}
\normalsize

\end{proof}

\subsection{Proof of Theorem 3}
\begin{proof}From the results of Theorem 2, it is straightforward to show that the order of the mean squared error of $\hat m(x)$ is $h^4+(T^2h^2)^{-1}+h^{*4}+T^{-1}+(nTh^*)^{-1}$. To minimize the mean squared error, we can taken $h=O(T^{-1/3})$ and $h^*=O((nT)^{-1/5})$.  Under such choices of $h$ and $h^*$, the order of the mean squared error is $(nT)^{-4/5}+T^{-1}$.

Therefore, to match the optimal nonparametric convergence rate $(nT)^{-4/5}$ for mean squared error, the condition $n=O(T^{1/4})$ is required. 
\end{proof}

\subsection{Proof of Theorem 4}
\begin{proof}

We start from the asymptotic expansion  from \eqref{eq::hat_beta_expansion} in the proof of Theorem 2. First, we investigate the asymptotic behavior of the third term on the right hand side of \eqref{eq::hat_beta_expansion}.

As a first step, we have 
\begin{align}
\Var\left(\sum_{t=1}^T(U_1(x_{it})-\bar U_1(x_{i\cdot}))^2\right)\leq 8 E\left(\left[\sum_{t=1}^T U_1(x_{it})^2\right]^2\right).
\end{align}

Now, following the definition of $U_1(\cdot)$ and applying the same expansion of $\omega_{1s}(x_{it})$ as in the proof of Theorem 1,
\begin{align*}
&E\left(\left[\sum_{t=1}^T U_1(x_{it})^2\right]^2\right)\\
=& \text{E}\biggm(\bigg[\sum_{t=1}^T\Bigm[\frac{\sum_{s=1}^TK_h(x_{it}-x_{1s})\sigma_1(x_{1s})\epsilon_{1s}}{Tf(x_{it})}\Bigm]^2\bigg]^2\biggm)(1+o(1))\\
\leq & \frac{1}{T^4}\text{E}\biggm(\sum_{s,u=1}^T\Big(\sum_{t=1}^T\frac{K_h^2(x_{it}-x_{1s})}{f^2(x_{it})}\text{I}_{\{|x_{1s}-x_{1u}|<2h\}}\Big)^2\sigma^2_1(x_{1s})\sigma^2_1(x_{1u})
\biggm)(1+o(1)),
\end{align*}
where the last inequality follows from exchanging the summation order and the property of the kernel function $K(\cdot)$. Observe that $f(\cdot)$ is bounded from below by Condition C2, the following inequality sequence is obtained. 
\begin{align*}
&E\left(\left[\sum_{t=1}^T U_1(x_{it})^2\right]^2\right) \\
\leq &\frac{1}{T^4\delta^4}\text{E}\biggm(\sum_{s,u=1}^T\Big( \frac{T\nu_0f(x_{1s})}{h}\Big)^2\text{I}_{\{|x_{1s}-x_{1u}|<2h\}}\sigma^2_1(x_{1s})\sigma^2_1(x_{1u})\biggm)(1+o(1))\\
\leq&\frac{O(1)}{T^2h^2}\sum_{s,u=1}^T\text{I}_{\{|x_{1s}-x_{1u}|<2h\}},
\end{align*}
where the last term has the order of $O(h^{-1})$ by noticing 
$$\sum_{s,u=1}^T\text{I}_{\{|x_{1s}-x_{1u}|<2h\}}=\sum_{s=1}^T4Thf(x_{1s}).$$
We can also derive the order of the variance for the other two terms, $$\Var\left(-\beta_i\frac{\sum_{t=1}^TW_{it}U_1(x_{it})}{\sum_{t=1}^TW_{it}^2}+\frac{\sum_{t=1}^TW_{it}\sigma_i(x_{it})\epsilon_{it}}{\sum_{t=1}^TW_{it}^2}\right)=O(T^{-1}).$$
 Due to the relationship of $h$ and $T$, the third term is negligible when calculating the asymptotic variance. Then,  the expansion for the bias of $\hat\beta_i$ can be rewritten as follows
\begin{align*}
&\hat\beta_i-\beta_i+\beta_i(h^2P_i+\frac{1}{Th}Q_i)\\
=&  -\beta_i\sum_{s=1}^T\left(\frac{W_{1s}\sigma_1(x_{1s})\epsilon_{1s}}{\sum_{t=1}^TW_{it}^2}\right) +\frac{\sum_{t=1}^T{W_{it}\sigma_i(x_{it})\epsilon_{it}}}{\sum_{t=1}^TW_{it}^2}(1+o_p(1))+o\Big(h^2+\frac{1}{Th}\Big),
\end{align*}
where the right hand side is an independent sum of random variables 
with their variances being of the same order, $O(T^{-1})$.
As a result, the central limit theorem can be applied directly for $\hat\beta_i$.
$$\sqrt T [\hat\beta_i-\beta_i-\beta_i(h^2P_i+\frac{1}{Th}Q_i)]\rightarrow^d N(0,\sigma_i^{*2}), $$
where the asymptotic variance $\sigma_i^{*2}$ is finite with the following expression.\begin{align*}
\sigma^{*2}&=\lim_{T\rightarrow\infty}\left[\frac{T^{-1}\sum_{t=1}^TW_{it}^2\sigma_i^2(x_{it})}{(T^{-1}\sum_{t=1}^TW_{it}^2)^2}+\beta_i^2 \frac{T^{-1}\sum_{t=1}^TW_{1t}^2\sigma_1^2(x_{1t})}{(T^{-1}\sum_{t=1}^TW_{it}^2)^2}\right].\\
\end{align*}
Notice that if the order of $h$ is between $T^{-\frac{1}{2}}$ and $T^{-\frac{1}{4}}$, then $\hat\beta_i$ is asymptotic unbiased since $\sqrt T\beta_i(h^2P_i+\frac{1}{Th}Q_i)\rightarrow^d 0$.

From Theorems \ref{th1} and \ref{th2} we have $\hat\beta_i,\hat\alpha_i,\hat m(\cdot)$ are consistent estimators of $\beta_i,\alpha_i, m(\cdot)$, respectively. Thus, $\hat\sigma_i^2=\frac{1}{T}\sum_{t=1}^T(y_{it}-\hat\alpha_i-\hat\beta_i\hat m(x_{it}))^2$ is also consistent for the variance under the assumption that $\sigma_i(\cdot)$ is a constant function for each subject $i$.
\end{proof}

\end{document}